\documentstyle[epsf]{article}

\newcommand{\newsection}{ \setcounter{equation}{0} \section}
\newcommand{\beq}{\begin{equation}} \newcommand{\eeq}{\end{equation}}
\newcommand{\bea}{\begin{eqnarray}} \newcommand{\eea}{\end{eqnarray}}

  \newcommand
{\Romannumeral}[1]{\uppercase\expandafter{\romannumeral#1}}

\newcommand{\be}{\begin{enumerate}} \newcommand{\ee}{\end{enumerate}}
\newcommand{\bi}{\begin{itemize}} \newcommand{\ei}{\end{itemize}}
\newcommand{\ba}{\begin{array}} \newcommand{\ea}{\end{array}}
\newcommand{\bc}{\begin{center}} \newcommand{\ec}{\end{center}}
\newcommand{\bt}{\begin{tabular}} \newcommand{\et}{\end{tabular}}

\def\lsim{\mathrel{\rlap{\lower4pt\hbox{\hskip1pt$\sim$}}
    \raise1pt\hbox{$<$}}}           
\def\gsim{\mathrel{\rlap{\lower4pt\hbox{\hskip1pt$\sim$}}
    \raise1pt\hbox{$>$}}}           

\newcommand{\Dslash}{{\hbox{D}\kern-0.6em\raise0.15ex\hbox{/}}} 

\begin{document}

\setlength{\oddsidemargin}{0cm} \setlength{\baselineskip}{7mm}

\input epsf




\begin{normalsize}\begin{flushright}
    UCI-98-31 \\
    September 1998 \\
\end{flushright}\end{normalsize}

\begin{center}
  
\vspace{50pt}
  
{\Large \bf {\it AENEAS} }

\vspace{20pt}

{\Large \bf A Custom-built Parallel Supercomputer }

{\Large \bf for }

{\Large \bf Quantum Gravity }

\vspace{40pt}
  
{\sl Herbert W. Hamber}
$^{}$\footnote{e-mail address : hhamber@uci.edu;
http://aeneas.ps.uci.edu/aeneas}
\\

\vspace{20pt}


Department of Physics and Astronomy \\
University of California \\
Irvine, CA 92171 \\

\end{center}

\vspace{40pt}

\begin{center} {\bf ABSTRACT } \end{center}
\vspace{12pt}
\noindent

Accurate Quantum Gravity calculations, based on the simplicial lattice
formulation, are computationally very demanding and require vast
amounts of computer resources.
A custom-made 64-node parallel supercomputer capable of performing up to
$2 \times 10^{10}$ floating point operations per second has been
assembled entirely out of commodity components, and has been
operational for the last ten months.
It will allow the numerical computation of a variety of quantities
of physical interest in quantum gravity and related field theories,
including the estimate of the critical exponents in the vicinity of
the ultraviolet fixed point to an accuracy of a few percent.

\vspace{24pt}

\vfill

\newpage

\vskip 10pt
\newsection{Physics Goals}
\hspace*{\parindent}

One of the outstanding problems in theoretical physics is
the determination of the quantum-mechanical properties
of Eintein's relativistic theory of Gravitation.
Approaches based on linearized perturbation methods have not been
successful so far,
as the underlying theory is known not to be perturbatively renormalizable.
Due to the complexity of even such approximate calculations, 
a fundamental coupling of the theory, the bare cosmological constant
term is usually set to zero, thus further reducing the potential physical
relevance of the results.
Furthermore, gravitational fields are themselves the source for
gravitation,
which leads to the problem of a highly non-linear theory, where
any sort of perturbative results is possibly of doubtful validity,
especially in the quantum domain, where strong fluctuations in
the gravitational fields appear at short distances.

The above-described situation bears some resemblance to the theory
of strong interactions, Quantum Chromodynamics. Non-linear
effects are known here to play an important role, and end up restricting
the validity of
perturbative calculations to the high energy, short distance regime
where the effective coupling can be considered weak.
For low energy properties Wilson's discrete lattice formulation, combined with
computer simulations, has provided so far the only convincing
evidence for quark confinement and chiral symmetry breaking,
two phenomena which are completely invisible to any order in the
weak coupling, perturbative expansion.

A discrete lattice formulation can
be applied to the problem of quantizing gravity as well. 
Instead of continuous fields, one deals with gravitational
fields which live only on discrete space-time points and interact
locally with each other.
In Regge's simplicial formulation of gravity ~\cite{re} one approximates
the functional integration over continuous metrics by a discretized
sum over piecewise linear simplicial geometries ~\cite{rw,le}.
In such a model the role of the continuum metric is played
by the edge lengths of the simplices, while curvature is
described by a set of deficit angles, which can be computed via
standard formulae as functions of the given edge lengths.
The simplicial lattice
formulation of gravity is locally gauge invariant ~\cite{ga}, and is known
to contain perturbative gravitons in the lattice weak field expansion,
making it an attractive lattice regularization of the continuum
theory.

The starting point for a non-perturbative study of quantum gravity
is a suitable definition of the discrete Feynman path integral.
In the simplicial lattice approach one starts from the discretized
Euclidean path integral for pure gravity,
with the squared edge lengths as fundamental variables,
\beq
Z_L \; = \; \int_0^\infty \; \prod_s \; \left ( V_d (s) \right )^{\sigma} \;
\prod_{ ij } \, dl_{ij}^2 \; \Theta [l_{ij}^2]  \; 
\exp \left \{ 
- \sum_h \, \Bigl ( \lambda \, V_h - k \, \delta_h A_h 
+ a \, { \delta_h^2 A_h^2 \over V_h } + \cdots \Bigr ) \right \}  \;\; .
\label{eq:zlatt} 
\eeq
The above expression represents an elegant discretization of the
continuum Euclidean path integral for pure quantum gravity
\beq
Z_C \; = \; \int \prod_x \;
\left ( {\textstyle \sqrt{g(x)} \displaystyle} \right )^{\sigma}
\; \prod_{ \mu \ge \nu } \, d g_{ \mu \nu } (x) \;
\exp \left \{ 
- \int d^4 x \, \sqrt g \, \Bigl ( \lambda - { k \over 2 } \, R
+ { a \over 4 } \, R_{\mu\nu\rho\sigma} R^{\mu\nu\rho\sigma}
+ \cdots \Bigr ) \right \}  \;\; ,
\label{eq:zcont}
\eeq
with $k^{-1} = 8 \pi G $, and $G$ Newton's constant, and reduces
to it for smooth enough field configurations.
In the discrete case the integration over metrics is replaced by
integrals over the elementary lattice degrees of freedom,
the squared edge lengths. The above discrete gravitational measure
is then the lattice analog of the DeWitt continuum functional
measure ~\cite{cms,ha}.
The $\delta A$ term in the lattice action is the well-known Regge
term, and reduces to the Einstein-Hilbert action
in the lattice continuum limit.
A cosmological constant term is needed for convergence of the path
integral, while the curvature squared term allows one to control the
fluctuations in the curvature. In practice, and for phenomenological
reasons, one is only interested
in the limit when the higher derivative terms are small,
$a \rightarrow 0$.
In this limit the theory depends, in the
absence of matter and after a suitable rescaling of the metric, only on
one bare parameter, the dimensionless coupling $k^2 / \lambda $.

The discretized theory contains only a finite set of variables, once
a set of suitable boundary conditions are imposed, such as open or
periodic.
In the end the original continuum theory of gravity is to be recovered
as the space-time volume is made large and the fundamental lattice
spacing  of the discrete theory is sent to zero, possibly without
having to rely, at least in principle, on any further approximation
to the original continuum theory.

Quantum fluctuations in the underlying geometry are represented
in the discrete theory
by fluctuations in the edge lengths, which can be modeled by
a well-defined Monte-Carlo stochastic process.
In analogy with other field theory models studied by computer,
all calculations so far have been performed in the Euclidean,
imaginary time framework, which is the only formulation amenable
to a controlled numerical study, at least for the foreseeable future.
The Monte-Carlo method, based on the concept of importance sampling,
is well suited for evaluating the discrete path integral for
gravity and for computing the required averages and correlation
functions.
By a careful analysis of the lattice results, the critical exponents
can be extracted, and the scaling properties of invariant correlation
functions determined from first principles.

Studies on small lattices suggest a rich scenario
for the gound state of quantum gravity ~\cite{ph}.
The present evidence suggests that simplicial quantum gravity
in four dimensions exhibits a phase transition (in $G$) between two phases:
a strong coupling phase, in which the geometry is smooth at large
scales and quantum fluctuations in the gravitational field are
bounded, and a weak coupling phase, in which the geometry
is degenerate and space-time collapses into a lower-dimensional
manifold.
Only the smooth, small negative curvature $AdS$ phase appears to
be physically acceptable.
The existence of a phase transition at finite coupling $G$ implies
non-trivial, calculable non-perturbative scaling properties for the
correlations and coupling constants of the theory, and in particular
Newton's constant.
Its presence is usually inferred from the presence of non-analytic terms
in invariant averages, such as for example the average curvature
\beq
<l^2> { < \int d^4 x \, \sqrt{ g } \, R(x) >
\over < \int d^4 x \, \sqrt{ g } > }
\; \equiv \; {\cal R} (k)
\mathrel{\mathop\sim_{ k \rightarrow k_c}}
- A_{\cal R} \; ( k_c - k ) ^{ 4 \nu - 1 } \;\; ,
\label{eq:rk}
\eeq
From such averages one can determine the value for $\nu$,
the correlation length exponent,
\beq
\xi (k) \; \mathrel{\mathop\sim_{ k \rightarrow k_c}} \; A_\xi \;
( k_c - k ) ^{ -\nu } \;\; .
\label{eq:mk}
\eeq
An equivalent result, relating the quantum expectation value
of the curvature to the physical correlation length $\xi$ , is
\beq
{\cal R} ( \xi ) \; \mathrel{\mathop\sim_{ k \rightarrow k_c}} \;
\xi^{ 1 / \nu - 4 }
\;\; .
\label{eq:rm}
\eeq
Matching of dimensionalities in these equations is restored by
supplying appropriate powers of the Planck length $\sqrt{G}$.
The exponent $\nu$ is known to be related to the derivative of the beta
function for $G$ in the vicinity of the ultraviolet fixed point,
\beq
\beta ' (G_c) \, = \, - 1/ \nu \;\; .
\eeq
In addition, the correlation length $\xi$ itself determines the
long-distance decay of the connected, invariant correlations at
fixed geodesic distance $d$.
Thus for the curvature correlation one has at large distances
\beq
< \sqrt{g} \; R(x) \; \sqrt{g} \; R(y) \; \delta ( | x - y | -d ) >_c \;
\mathrel{\mathop\sim_{d \; \gg \; \xi }} \;\;
d^{- \sigma } \; e^{-d / \xi } \;\; ,
\label{eq:exp}
\eeq
while for shorter distances one expects a slower power law decay
\beq
< \sqrt{g} \; R(x) \; \sqrt{g} \; R(y) \; \delta ( | x - y | -d ) >_c \;
\mathrel{\mathop\sim_{d \; \ll \; \xi }} \;\; 
{1 \over d^{\; 2(4-1/ \nu)} } \;\; .
\label{eq:pow}
\eeq
The scale dependence of the effective Newton constant is given by
\beq
G(r) \; = \; G(0) \left [ 1 \, + \, c \, ( r / \xi )^{1 / \nu} \, 
+ \, O (( r / \xi )^{2 / \nu} ) \right ] \;\; ,
\eeq
with $c$ a calculable numerical constant.
In this last expression the momentum scale $\xi^{-1}$ plays a role
somewhat similar to the
scaling violation parameter $\Lambda_{\overline{MS}}$ of QCD.
Hopefully it should be clear, even from this brief discussion, that the
critical exponents by themselves already provide a significant
amount of useful information about the continuum theory.

In reality, the complexity of the interaction and the need to sample
many statistically independent field configurations in the
path integral, which is necessary for correctly incorporating
into the model the effects of quantum-mechanical
fluctuations, leads to the requirement of powerful computational
resources.

\vskip 30pt
\section{The Machine}
\hspace*{\parindent}

In this section tha main physical characteristics of the machine will
be described.
Although the Aeneas computer was built with quantum field
theory applications in mind, it is in fact designed as
a general purpose supercomputer and therefore suitable
for a wide range of computationally intense applications.
Its layout is similar to other parallel distributed memory
architectures, such as the IBM SP2, the Intel Paragon or
the Cray T3E, although the switching network is not 
custom made and therefore not as fast.
To keep the costs under control and achieve a high
performance over cost ratio, it has been assembled out
of relatively easy available hardware and software
components.
The essential features of the machine are

\begin{itemize}

\item
Use of commodity microprocessors and data buses, allowing the use
of reliable and cost-effective off-the-shelf technology,

\item
Fast Ethernet (100 MegaBits/sec) connectivity between processor nodes,

\item
Reliable, well tested public domain software, based on Linux for the
operating system, on public domain Gnu tools and 
compilers, and on MPIch and PVM for inter-processor communication.

\end{itemize}

Before AENEAS was assembled, the above combination of commodity
hardware and public domain software
tools had been tested on similar machines based on
Pentium Pro processors built at NASA GSFC,
CalTech and LANL. The present design of Aeneas differs from these
earlier prototypes in that it uses faster 300 MHz PII processors,
newer design motherboards and larger, faster Ethernet switches.

In more detail, the hardware consists of 65 nodes, each based on

\begin{itemize}

\item
an Intel Pentium II processor with 512k cache running at 300 MHz

\item
a PCI motherboard based on the Intel 440LX chipset

\item
128 MegaBytes of 168-pin 10-ns SDRAM

\item
one 3.2 GBytes Quantum ST E-IDE disk drive

\item
one D-Link Fast Ethernet adapter with the DEC Tulip 21142 chipset

\item
a floppy drive and an S3 Virge video card for maintenance access

\end{itemize}

Multiple ethernet cards are installed in the front-end node in
order to provide external access to the cluster, isolating at the
same time the internal network traffic from the outside office network.
Alternatives to Fast Ethernet, such as Myrnet, offer slightly higher
performance, but at a much higher price per node. Given that a major
bottleneck in the design is the speed of the PCI bus (66 MHz), it
seemed unwise to afford the extra cost without reaping significant
performance improvement benefits.
Faster processors (such as the DEC alpha) would
improve single node performance (especially on long vector 
operations), but would still face the networking bottlenecks for
communication-intensive parallel applications.
Fast (pio-4) IDE disks where chosen because of their lower costs in
comparison to SCSI disks, which offer higher performance but at a
significantly higher price. Reliability is comparable for both types
of disks, as the disk and heads employ similar technology.
Several nodes, including the front end, have additional larger
6.4 GigaBytes IDE disks, to ease the movement of large data sets in and
out of the parallel machine.
Three spare nodes, as well as additional cloned spare disks, ensure
brief downtimes (of about 20 minutes or less) in case of a
hardware failure.

Connectivity between the processor nodes is provided by
two 36-port 3-Com Fast Ethernet (100 MegaBits/sec) crossbar switches,
which are connected to the processor nodes via standard cat5 twisted
pair cables.
At full duplex, any two nodes can exchange data with each other over the
switch at up to 25 MegaBytes/sec, the peak hardware bandwidth.
The Fast Ethernet crossbar switches themselves have a
backplane switching capability of  6.6 GigaBits/sec, and
are connected to each other via multiple trunked full duplex
Gigabit Ethernet SC multi-mode fiber connections (using the available
expansion modules), giving a peak hardware
bandwith between the two switches of 500 MegaBytes/sec.

\vskip 80pt

\begin{center}
\leavevmode
\epsfysize=7cm
\epsfbox{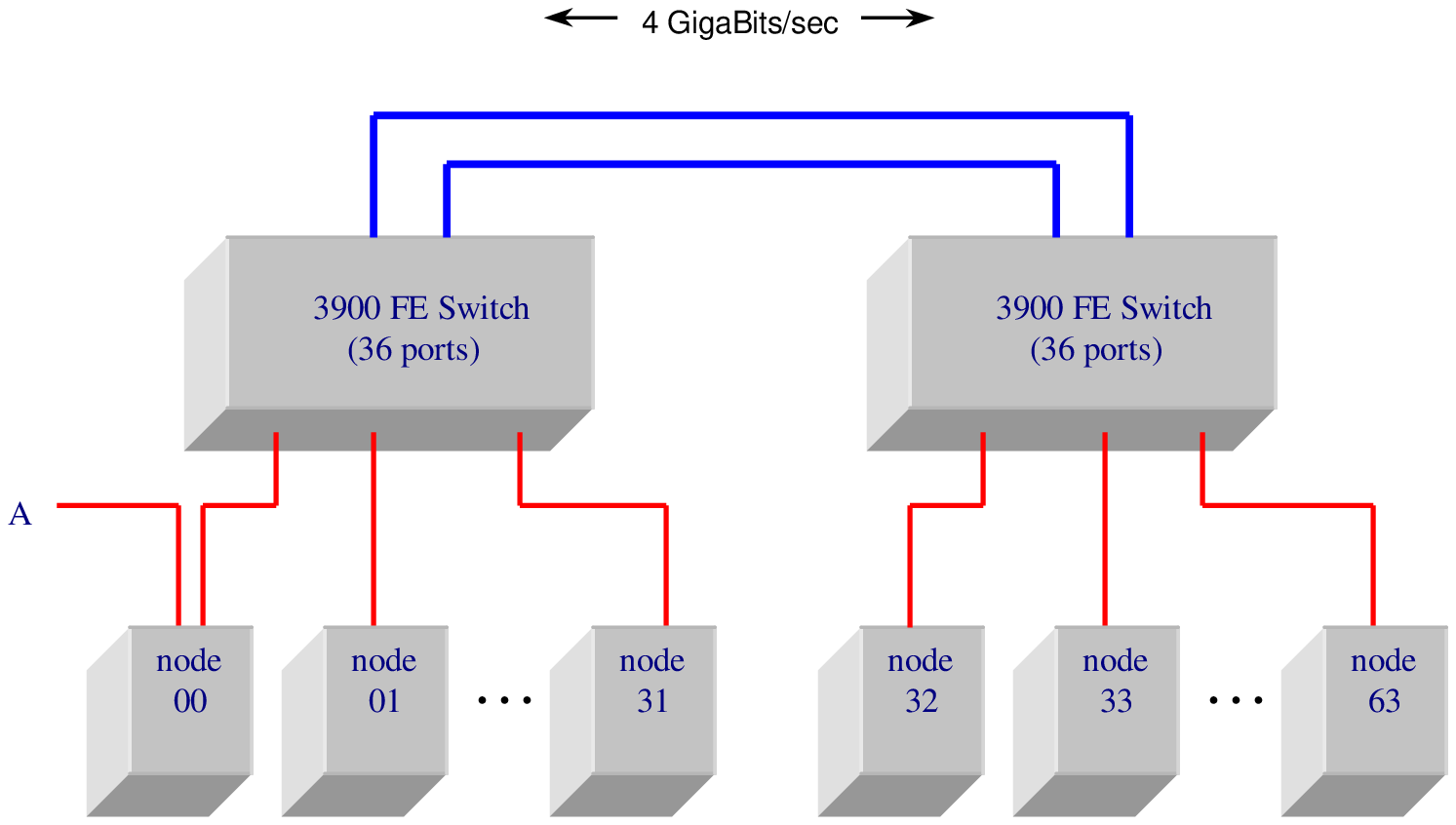}
\end{center}

\begin{center}
\center{\small {\it 
Network topology of the Aeneas parallel supercomputer.
\medskip}}
\end{center}

Power consumption is about 50 Watts/node.
As a significant amount of heat is dissipated by the 64 nodes,
the machine is housed in a machine room with a large chilled
water cooling unit, ensuring a constant $72^0F$ temperature
throughout the room.
The total cost of the hardware, including tax, is about
ninetytwothousand dollars at September 1998 market prices.

The software installed on all nodes includes

\begin{itemize}

\item
the RedHat Linux distribution, with updated Ethernet drivers

\item
Gnu g77, C, C++ compilers, and Absoft's f77 and f90 compilers

\item
MPI and PVM for message passing between processor nodes

\item
additional libraries such as CernLib and Lapack

\item
a queueing system based on DQS

\end{itemize}

The RedHat Linux distribution was chosen because of its ease
of installation and maintenance via rpm technology.
The software on all disks is cloned, providing for the same
user interface, compilers, libraries etc. on all nodes.
The latest releases of Linux (RH4.2-5.1) are very stable, and
the present cluster has registered uptimes of the order of months
on all nodes, which are clock synchronized via an xntp time server.
Both g77 and Absoft's f77 and f90 compilers work well with the
MPI distribution from Argonne National Lab (MPIch), with the
Absoft f77 and f90 compilers typically generating faster code than g77.

\vskip 30pt
\section{Performance}
\hspace*{\parindent}

In this section various performance aspects of the Aeneas
machine on a variety of codes will be discussed.

Single node code performance depends on a variety of factors,
including array sizes, cache size compatibility,
quality of coding, version of libraries used and quality of compiler
optimization. It is not unusual for typical Fortran code
to achieve floating point performance between 60 and 80 MegaFlops
on one node of Aeneas.
For long vector operations the MegaFlop rate rises to anywhere
between 90 and 250, with judicious use of unrolling and an
appropriate use of compiler switches. Typically the Pentium II at
300 MHz is about $50\%$
faster than the 200 MHz Pentium Pro (i.e. close to the ratio
of clock speeds), whit Absoft's f77 compiler typically
generating faster code (by as much as $32\%$) than g77.
The standard Linpack benchmark on a 100 by 100 matrix
multiply gives 85 MegaFlops in single precision, versus 66 MegaFlops
in double precision, a $22\%$ degradation in speed.
In general, the Pentium II's performance versus the Mips R10000,
the IBM P2SC and the Dec Alpha processors is quite good for
scalar-dominated code, but the Intel processor is somewhat
out-classed by code which can take advantage of the multiple
arithmetic units available on the other three processors.

Most parallel applications developed so far embed MPI (Message Passing
Interface, version 1.1, from Argonne National Lab) library calls in
f77 or f90 programs, although there is some limited PVM (Parallel
Virtual Machine, from Oak Ridge National Lab) parallel
programming experience.
In general MPI is preferred because of lower latencies and
higher bandwidth, and therefore higher parallel performance.
Most MPI codes have been ported
from larger parallel machines, such as the Cray T3-E 1200 and the IBM
SP2 (which use their own tuned MPI implementation).
As the different MPI versions
are virtually indistinguishable from one machine to another, very
few code modifications are necessary. These mostly concern
floating point precision issues (the ix86 processors are 32-bit chips)
and calls to the real time clock. In general, MPI with either f77
or f90 appears to work flawlessly under Linux.

\vskip 60pt

\begin{center}
\leavevmode
\epsfysize=7cm
\epsfbox{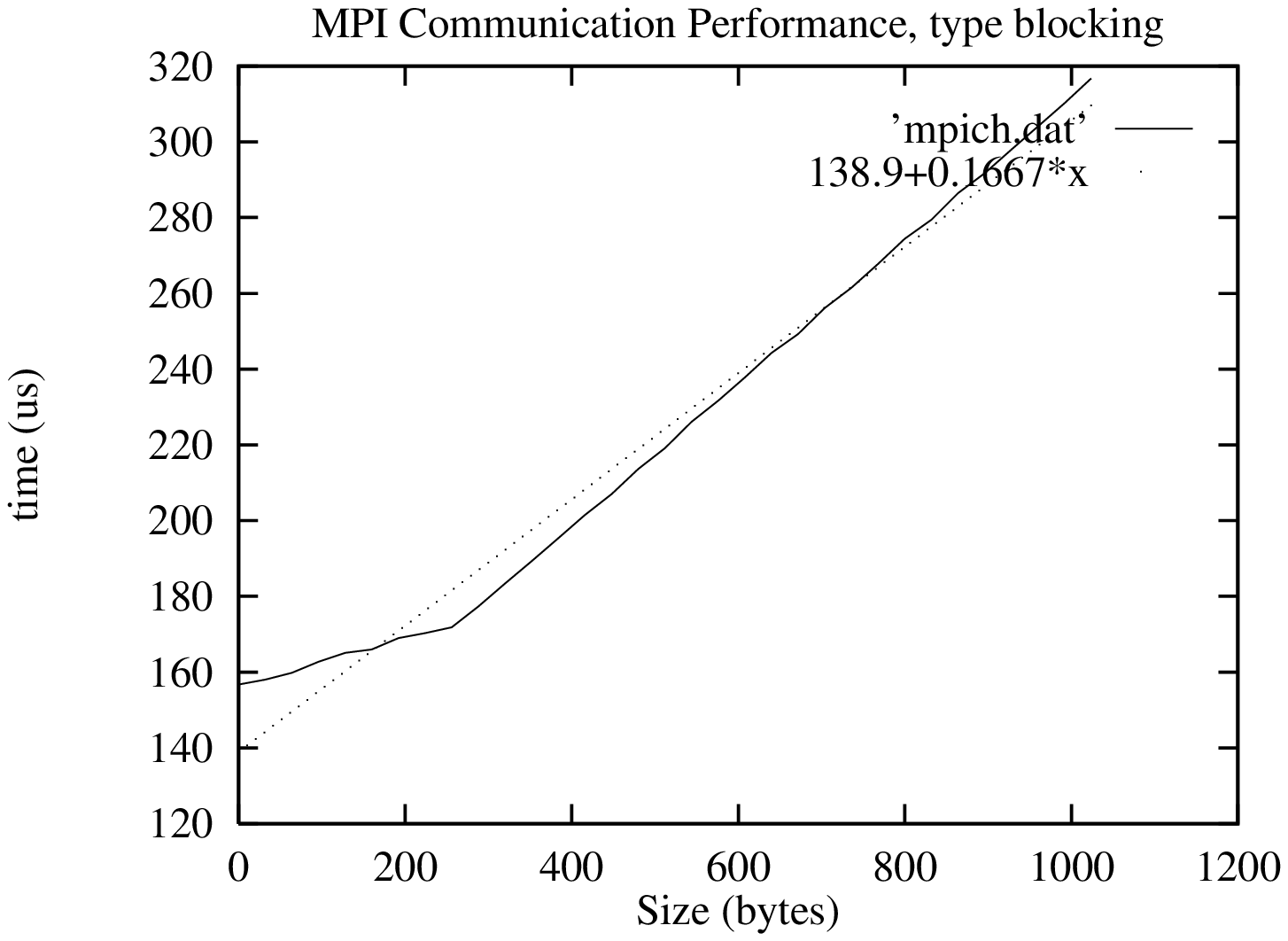}
\end{center}

\begin{center}
\center{\small {\it 
Communications performance benchmark, from MPIch distribution.
\medskip}}
\end{center}

It seems useful to compare the communication parameters of
Aeneas to other parallel supercomputers.
The standard MPI distribution from ANL includes a set of test programs
to evaluate the efficiency of parallel applications under
MPI. Tools are included to estimate both bandwith and latency
in interprocessor communications. Between any two nodes one finds
that the measured latency is about 150 $\mu secs$, while the measured
communication bandwidth ranges between 0.38 MegaBytes/sec
(for 64-byte packets) and 10.0 MegaBytes/sec (for 256k-byte packets).
The latter transfer rate figure is close to the peak one-way
hardware bandwith limit of 12.5 MegaBytes/sec.
The lesson here is that in general it pays to send one large packet
between nodes, as opposed to several small ones.
The above communications performance figures should be compared for
example to an IBM SP2, for which the bandwith is 120 MegaBytes/sec
and the latency 52 $\mu sec$.

A number of standard parallel benchmarks have been run, with the
intent of comparing Aeneas with similar parallel
computers available commercially, and asses its viability
as a machine for various problems involving quantum field theory simulations.
For such a comparison, a particularly popular suite is the Nasa NAS
(Numerical Aerodynamics Simulations)
class B  v. 2.3 set of parallel benchmarks, most of
which consist of various Navier-Stokes equation solver fragments, used
mostly in Computational Fluid Dynamics (CDF) applications.
Some of them involve a fair amount of linear algebra distributed
across the processors.
The NAS benchmarks clearly show the effectiveness of Aeneas as a
general-purpose parallel
supercomputer, compared to larger and much more costly machines
(a 64-node Cray Origin 2000, for example, has a list price of about \$1.9M).
Shown above, as an example, is a performance comparison for the LU
matrix decomposition benchmark with the latest Cray Origin 2000
(one should perhaps add that none of the quoted
NAS benchmarks have been tuned for the Aeneas architecture).
When a large number of nodes is used, the performance
degradation in the NAS benchmarks becomes more significant due
to the limitations in the network speed (communications overhead
for $N$ nodes generally grows like $N \log_2 N$).
Not unexpectedly, Aenas seems to be doing well performance-wise
when there is not an excessive amount of communication between
the nodes.

\vskip 60pt

\begin{center}
\leavevmode
\epsfysize=7cm
\epsfbox{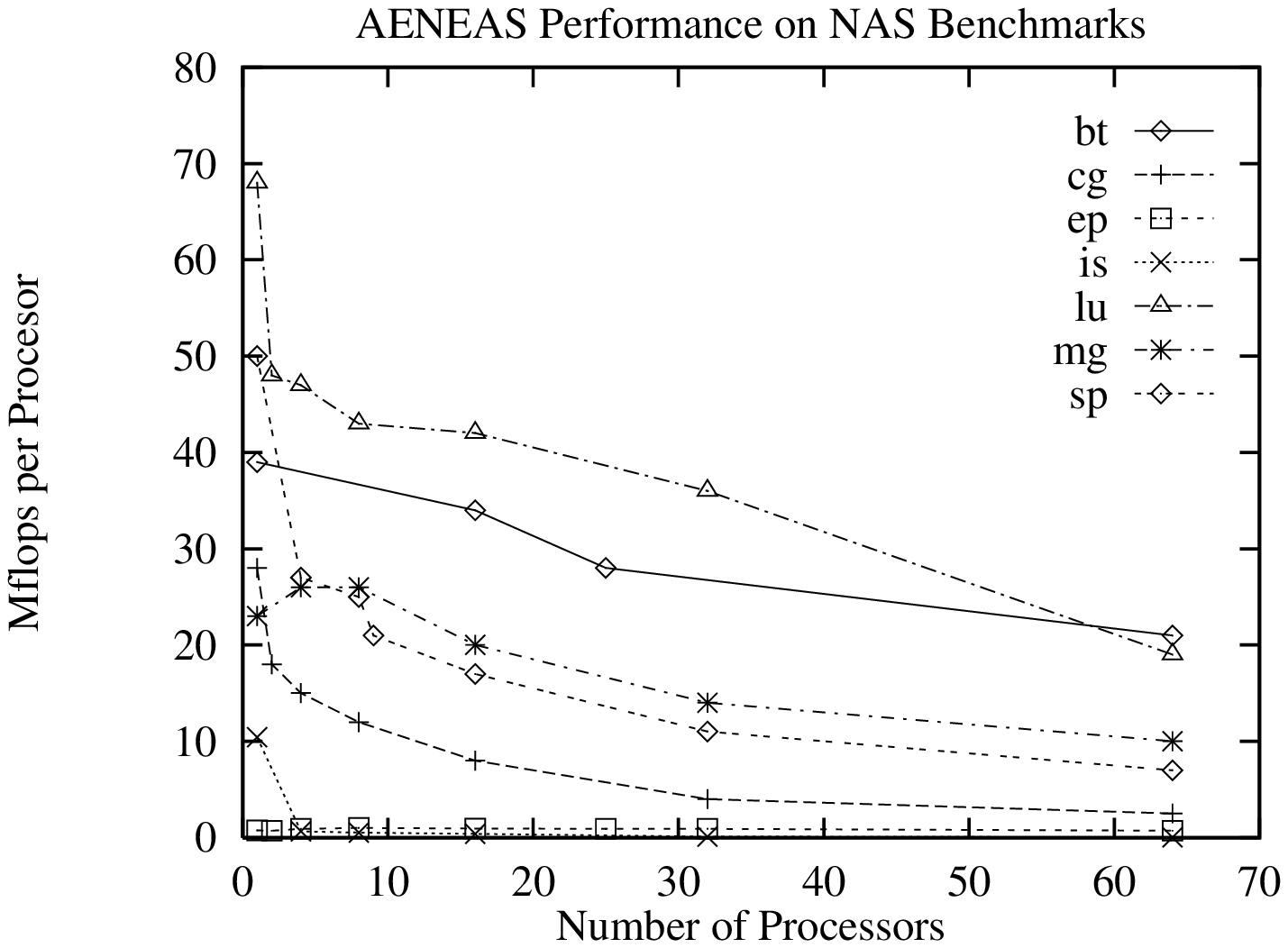}
\end{center}

\begin{center}
\center{\small {\it 
Scaling of NAS parallel benchmarks with processor nodes.
\medskip}}
\end{center}

\begin{table}

\begin{center}
\begin{tabular}{|l|l|l|l|l|}
\hline
  & 4 nodes & 8 nodes & 16 nodes & 32 nodes
\\ \hline \hline
Aeneas (300 MHz) & 187 & 346 & 675 & 1142 
\\ \hline
Cray Origin 2000 (195 MHz) & 367 & 696 & 1383 & 2991 
\\ \hline \hline
\end{tabular}
\end{center}
\label{lu}
\center{\small {\it
Nasa NAS Class B v2.3 parallel benchmark, LU decomposition.
Rates in MegaFlops.
\medskip}}


\end{table}

As a final example, the performance of the simplicial quantum gravity
code is shown in comparison to other similar parallel machines.
As will become clear below, the gravity application is well suited
for a parallel architecture, and
was in fact originally developed for the CM5, whose message passing
calls were rather similar to MPI. Later it was adapted slightly to
other architectures such as the IBM SP2 and the Cray T3E.
The particular example shown in the table is
for a lattice of $16^4$ sites, although much larger lattices 
such as $32^4$ are feasible.
These types of simulation codes are well suited for Aeneas,
as they involve relatively little communication between
processors versus actual computation. Typically the update
of one edge length requires of the order of $100k$ floating
point operations, versus the exchange of only 8 Bytes of data 
between nodes after the update has been completed.
Either 32 or 64 edges can be updated in parallel.
On a lattice with $16^4$ sites one can therefore typically generate
up to about $100k$ metric configurations a month.
The actual accuracy with which critical exponents and correlations
can be determined depends of course on a number of factors, of which
the lattice size and numerical accuracy are just two components.

The overall versatility of the machine is also brought out by the fact that
a number of applications in disciplines outside of theoretical
physics can use such an architecture for productive work.
Additional parallel applications which have been deployed successfully on
Aeneas include an atmospheric chemistry code for simulating the
diffusion of pollutants in the atmosphere, and a fluid dynamics
application involving turbine blade design.
In addition, the machine has been used effectively as a high-powered
cluster for high energy physics detector simulations and data
analysis, as part of the Amanda and Super-Kamiokande experimental projects.

\begin{table}

\begin{center}
\begin{tabular}{|l|l|l|l|}
\hline
  & 1 node & 16 nodes & 32 nodes
\\ \hline \hline
TMC CM5 & 36340 ({\bf 2}) & 2271 ({\bf 2}) & 1136 ({\bf 2})
\\ \hline
IBM SP2 (160 MHz) & 873 ({\bf 95}) & 69 ({\bf 75}) & 36 ({\bf 73}) 
\\ \hline
Aeneas (300 MHz) & 1090 ({\bf 77}) & 83 ({\bf 63}) & 48 ({\bf 55}) 
\\ \hline \hline
\end{tabular}
\end{center}
\label{grav}

\center{\small {\it
Performance of Simplicial Gravity code. Wall clock times in
seconds, MegaFlops per node in parenthesis.
\medskip}}


\end{table}
\vskip 10pt

\vspace{20pt}

{\bf Acknowledgements}

The author would like thank G. Veneziano and the Theoretical
Physics Division at CERN for hospitality.
In addition he would like to thank Don Becker of
Nasa Goddard and Jan Lindheim of CalTech for many useful informations
related to Linux PC clusters. 
The Aeneas Parallel Supercomputer project is supported by
the Department of Energy, the National Science Foundation and
the University of California.

\vspace{30pt}


\vfill
\newpage
\end{document}